\documentclass[12pt]{iopart}
\usepackage{geometry,graphicx, iopams,float,subfigure}

\expandafter\let\csname equation*\endcsname\relax

\expandafter\let\csname endequation*\endcsname\relax
\usepackage{amsmath}

\begin{document}

\title{Chaos in Bohmian Quantum Mechanics: A short review }

\author{ G. Contopoulos and A.C. Tzemos}

\address{Research Center for Astronomy and 
Applied Mathematics of the Academy of 
Athens - Soranou Efessiou 4, GR-11527 Athens, Greece}
\ead{gcontop@academyofathens.gr, thanasistzemos@gmail.com}
\vspace{10pt}

\begin{abstract}
This is a short review in the theory of chaos in Bohmian Quantum
Mechanics based on our series of works in this field.
Our first result is the development of a generic theoretical mechanism
responsible for the generation of chaos in an 
arbitrary Bohmian system (in 2 and 3 dimensions). This mechanism 
allows us to explore the effect of chaos on  Bohmian trajectories 
and study in detail (both analytically and numerically)  the different kinds of Bohmian trajectories where, in general, 
chaos and order coexist. Finally we explore the effect of quantum 
entanglement on the evolution of the Bohmian trajectories and study  chaos and ergodicity in qubit systems which are of great theoretical and 
practical interest. We find that the chaotic trajectories are also ergodic, i.e. they give the same final distribution of their points after a long time regardless of their initial conditions. In the case of strong entanglement most trajectories are chaotic and ergodic and an arbitrary initial distribution of particles will tends to Born's rule over the course of time. On the other hand, in the case of weak entanglement  the distribution of Born's rule is dominated by ordered trajectories and consequently  an arbitrary initial configuration of particles will not tend, in general, to  Born's rule, unless it is  initially satisfied. Our results shed light on a fundamental problem in Bohmian Mechanics, namely whether there is a dynamical approximation of Born's rule by an arbitrary initial distribution of
Bohmian particles.
\end{abstract}

\section{INTRODUCTION}
In Quantum Mechanics (QM) the state of a quantum particle described by a wavevector $|\Psi\rangle$ evolves according to the time-dependent Schr\"{o}dinger's equation which in the position representation reads:
\begin{align}
-\Big(\frac{\hbar^2}{2m}\nabla^2+V\Big)\Psi(\vec{r},t)=i\hbar\frac{\partial\Psi(\vec{r},t)}{\partial t},
\end{align}
where $V=V(\vec{r},t)$ is the potential, $\hbar$ is Planck's constant, $m$ is the mass and $\Psi=\langle \vec{r}|\Psi\rangle$ is the wavefunction corresponding to the state vector $|\Psi\rangle$. 
In the standard approach of QM (the Copenhagen approach, see, e.g.,\cite{ballentine1998quantum}) trajectories are not considered, because the Heisenberg uncertainty does not allow the simultaneous determination of positions and velocities.

Bohmian Quantum Mechanics (BQM) is an alternative interpretation of QM. It is a nonlocal pilot wave theory
whose  principles were first developed by Louis de Broglie \cite{debroglie1927a,debroglie1927b} and then by David Bohm \cite{Bohm, BohmII}.\footnote{BQM is also related to the hydrodynamical formulation of QM made by Madelung in 1927 \cite{madelung1927quantentheorie}.} According to BMQ, the quantum particles follow certain trajectories guided by the wavefunction according to a set of deterministic equations, the so called Bohmian equations:
\begin{align}
m\frac{d\vec{r}}{dt}=\hbar\Im\Big(\frac{\nabla\Psi}{\Psi}\Big)\label{bohmeq}
\end{align}
BQM \cite{ bell1987speakable, holland1995quantum} assumes that such trajectories exist, whether they are observed or not and for this reason it is called ``ontological approach''. However the two approaches are intimately linked and in most cases they predict the same experimental results.\footnote{There are only exceptions referring to the measurement of time, as the one that has has been considered in detail by Delis, Efthymiopoulos and Contopoulos in \cite{delis2012quantum}.}  In recent years there has been a remarkable interest in Bohmian Mechanics \cite{trahan2005quantum, nikolic2008, pladevall2012applied, sanz2012trajectory, sanz2013trajectory, sanz2019bohm}. Furthermore there have been experiments that use weak measurements in order to calculate average trajectories, in the double slit experiment, that are consistent with the theoretical Bohmian trajectories \cite{kocsis2011observing}.

In Classical Mechanics (CM), where the state of a system is completely known for given $(\vec{r},\vec{p})$, the equations of motion are given by second order differential equations
\begin{align}
m\frac{d^2\vec{r}}{dt^2}=-\nabla V,
\end{align}
that depend only on the potential $V$. 
In BQM a complete description of the quantum state is not given by $\Psi$ alone but by $(\Psi,\vec{r})$, namely by the wavefunction and the positions of the  particles.  In order to compare the classical and  quantum motions we write the Bohmian equations of motion in the form:
\begin{align}
m\frac{d^2\vec{r}}{dt^2}=-\nabla(V+Q),
\end{align}
where the quantity
\begin{align}
Q=-\frac{\hbar^2}{2m}\frac{\nabla^2|\Psi|}{|\Psi|}
\end{align}
is the so called ``quantum potential''.
An important  question now is the relation between the classical equations of motion and their quantum analogues.
There have been efforts to  approach the classical equations of motion from the quantum ones by considering $\hbar\to 0$, according to the so called ``correspondence principle''. However, in most cases,  the wavefunctions contain factors of the form $\exp(-x^2/2\hbar)$, and as the quantity $\nabla^2|\Psi|$ contains a factor $1/\hbar^2$, the quantum potential  does not tend to zero as $\hbar\to 0$. As a consequence the quantum trajectories are in general quite different from the classical trajectories.

A question that arises naturally from the study of the relation between QM and CM is whether we can observe chaos in QM.  In standard QM  several approaches have been employed in order to define and study quantum chaos since trajectories are not considered and the dynamics is governed by Schr\"{o}dinger's equation which is linear \cite{berry1987quantum, gutzwiller2013chaos,robnik2016fundamental}.  The usual approach in standard QM is to 
study the behaviour of quantum systems when the corresponding classical systems are chaotic.
 
On the other hand, the existence of trajectories in BQM allows us to define and study quantum chaotic behaviour 
by applying all the techniques of classical dynamical systems. Thus chaos in Bohmian Dynamics has attracted a lot of interest in  the last decades \cite{durr1992quantum, parmenter1995deterministic, 
faisal1995unified, schwengelbeck1995definition, iacomelli1996regular, sengupta1996quantum,  de1996exponential, 
parmenter1997chaotic,  frisk1997properties, konkel1998regular, wu1999quantum, cushing2000bohmian, makowski2000chaotic,  falsaperla2003motion,  wisniacki2005motion, wisniacki2006vortex,   wisniacki2007vortex, borondo2009dynamical,  sengupta2014bohmian,  cesa2016chaotic}.

As we will see in the next section, within the framework of BQM there are examples of ordered classical systems which are chaotic quantum mechanically and vice versa, chaotic classical systems which are ordered quantum mechanically. 
The results of the above study give us the opportunity to shed light on fundamental aspects of BQM, such as the dynamical approximation of Born's rule and the relation between chaos and entanglement.

In Section 2 of the present paper we make a first comparison between the behaviour of classical and quantum dynamical systems regarding chaos, using some characteristic physical systems. In Sections 3-5 we present the basic theoretical mechanism responsible for the emergence of chaos in 2-dimensional quantum systems. In Section 6 we 
present chaos in 3-dimensional quantum systems and describe a special case 
of 3d systems, the partially integrable systems, where the trajectories evolve on 2-d integral surfaces embedded in 3-d space. Then in Section 7 we present our main results for the interplay between chaos and entanglement in BQM and its implications on the dynamical approximation of Born's rule, by studying a simple two-qubit system. Finally in Section 8 we draw our conclusions.

\section{ORDERED AND CHAOTIC BOHMIAN TRAJECTORIES}

The Bohmian trajectories may be ordered or chaotic. The distinction is based on the value of the Lyapunov characteristic number (LCN)
\begin{equation}
LCN=\lim_{t\to\infty}(\chi),\quad \chi=\frac{1}{t}\ln\Big(\frac{|\xi|}{|\xi_0|}\Big),
\end{equation}
where $\chi$ is the ``finite time LCN'' and $\xi,\xi_0$ the infinitesimal deviations of nearby trajectories at times $t$ and $t_0$ (given by the variational equations)\cite{efthymiopoulos2006chaos}.

\begin{figure}[H]
\centering
\includegraphics[scale=0.3]{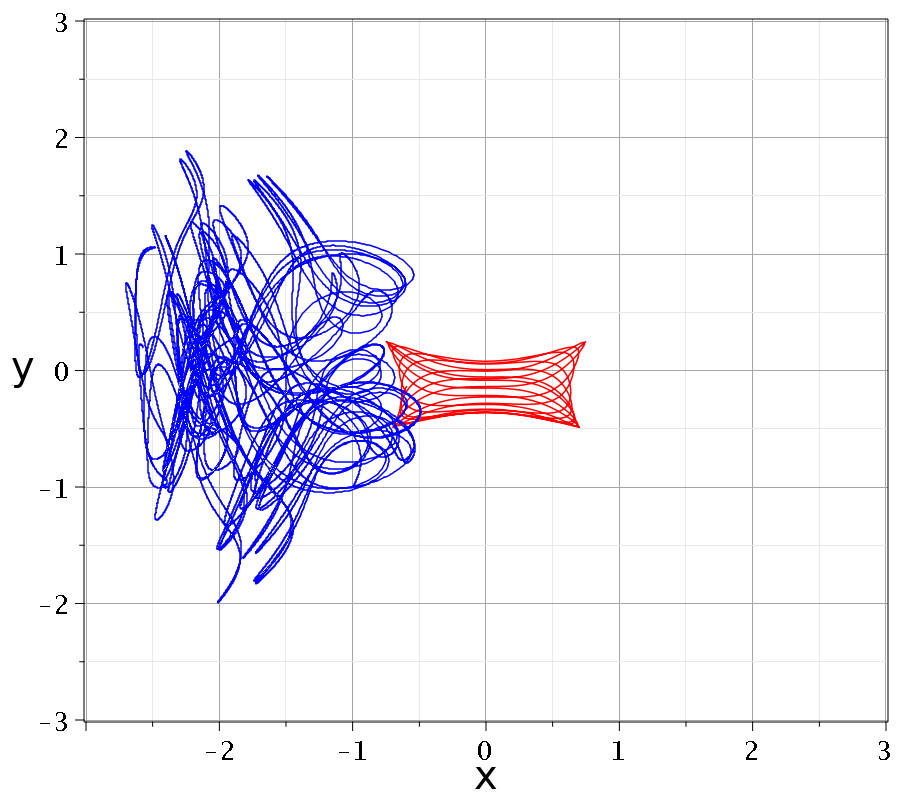}
\caption{An ordered trajectory (red) and a chaotic trajectory (blue) of the system \eqref{sys9} for  $\omega_1=1,\omega_2=1/\sqrt{2}, c_1=1/\sqrt{2}, c_2=1/2$. Initial conditions:$x_0=0.75, y_0=0.25$ (ordered) and $x_0=y_0=-1$ (chaotic).}\label{troxies}
\end{figure}

\begin{figure}[H]
\centering
\includegraphics[scale=0.3]{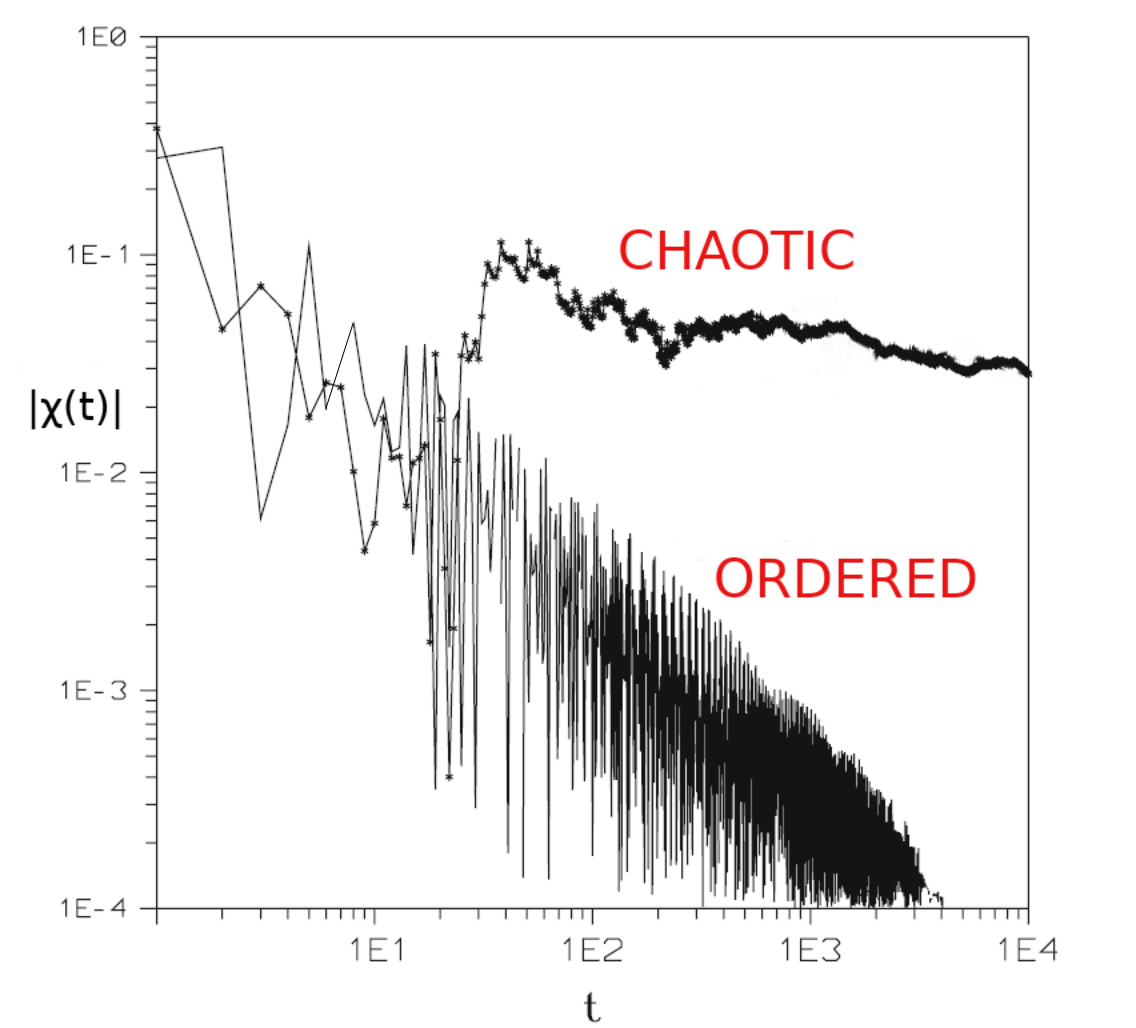}
\caption{The absolute value of the finite time LCN, $|\chi|$, for the trajectories of Fig.~\ref{troxies}.}.\label{LCN}
\end{figure}

We have found examples of ordered classical trajectories which are chaotic in Bohmian mechanics  and vice versa.
A first example is given by a system of 2 harmonic oscillators with Hamiltonian considered by Parmenter and Valentine in \cite{parmenter1995deterministic}
\begin{align}\label{hamil}
H=\frac{1}{2}(\dot{x}^2+\omega_1^2x^2+\dot{y}^2+\omega_2^2y^2)
\end{align}
which is completely integrable in classical Mechanics and all the trajectories are ordered (Lissajous figures). The  corresponding Schr\"{o}dinger's equation has solutions of the form:
\begin{align}
\Psi_{n_1,n_2}=\frac{(\omega_1\omega_2)^{\frac{1}{4}}}{\sqrt{2^{n_1+n_2}n_1!n_2!\pi}}\exp(-iE_{n_1,n_2}t)\exp\Big(-\frac{1}{2}(\omega_1x^2+\omega_2y^2)\Big)H_{n_1}(\sqrt{\omega_1}x)H_{n_2}(\sqrt{\omega_2}y),\label{wavefunction}
\end{align}
with $m=\hbar=1$, $E_{n_1,n_2}=\omega_1(1/2+n_1)+\omega_2(1/2+n_2)$, and $H_{n_1}, H_{n_2}$ are Hermite polynomials. If we take as $\Psi$,   a coherent superposition of three solutions of the form:
\begin{align}
\Psi=\Psi_{0,0}+c_1\Psi_{1,0}+c_2\Psi_{1,1},\label{sys9}
\end{align} then by use of the Bohmian equations of motion \eqref{bohmeq}, we find that in the case of irrational ratio $\omega_2/\omega_1$ there exist both ordered and chaotic trajectories, as shown in Fig.~\ref{troxies}.
If we calculate the LCN of these trajectories we find (Fig.~\ref{LCN}) that in the first case $|\chi|$ decreases in time inversely proportionally to $t$, while in the second case it saturates at a constant positive value after a transient time period. Therefore in the ordered case the LCN is zero, while in the chaotic case LCN it is a positive number.


On the other hand the Hamiltonian \cite{contopoulos2008order}
\begin{align}
H=\frac{1}{2}(\dot{x}^2+x^2)+\frac{\varepsilon x^4}{4}\exp\Big(-\frac{x^2}{2\sigma^2}\Big)\cos(\omega t)\label{Ham}
\end{align}
has appreciable classical chaos. In Fig.3 we show the distribution of the points $(x,\dot{x})$ of a trajectory whenever $t=k T=2k\pi/\omega$ with $k=1,2...10000.$ These are the intersections of the trajectory with a stroboscopic Poincar\'{e} surface of section. This distribution is mostly chaotic and, apparently, if we calculate the LCN of  trajectories in the chaotic domain, this is is clearly positive (Fig.~\ref{lcn}a). On the other hand the corresponding quantum equation of motion is of the  form
\begin{align}
\frac{dx}{dt}=f(x,t)
\end{align}
which has no chaos at all. In fact the LCN of the  corresponding quantum trajectory is zero. 

This difference is due to the fact that in the classical case Eq.~\eqref{Ham}  represents a Hamiltonian dynamical system of $1.5$ degrees of freedom,  which allows the existence of chaos,  while in  the quantum case we have a non-Hamiltonian system of 1.5 degrees of freedom, and thus cannot have chaotic trajectories. Thus the LCN can be  finite in the classical case (Fig.~\ref{lcn}a) while it is always zero in the quantum case (Fig.~\ref{lcn}b).

\begin{figure}[H]
\centering
\includegraphics[scale=0.3]{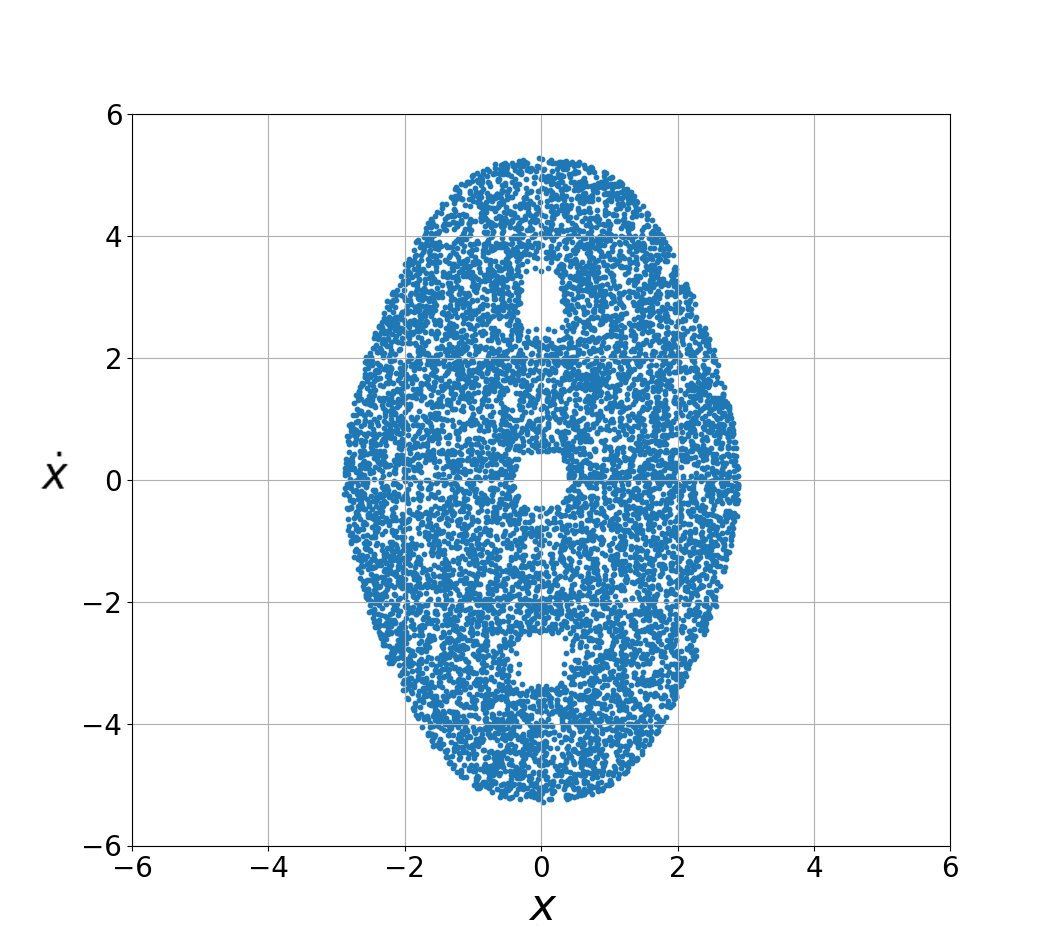}
\caption{The stroboscopic Poincar\'{e} suface of section of the Hamiltonian \eqref{Ham}. Distribution of the points of a trajectory with initial conditions $x(0)=1,  \dot{x}(0)=1$, when $t=2k\pi/\omega$ with $k=1\dots 10000$. }\label{stroboscopic}
\end{figure}

As a consequence the distinction between order and chaos in quantum mechanics requires the Bohmian calculations of trajectories.

\begin{figure}[H]
\centering
\includegraphics[width=7cm,height=6cm]{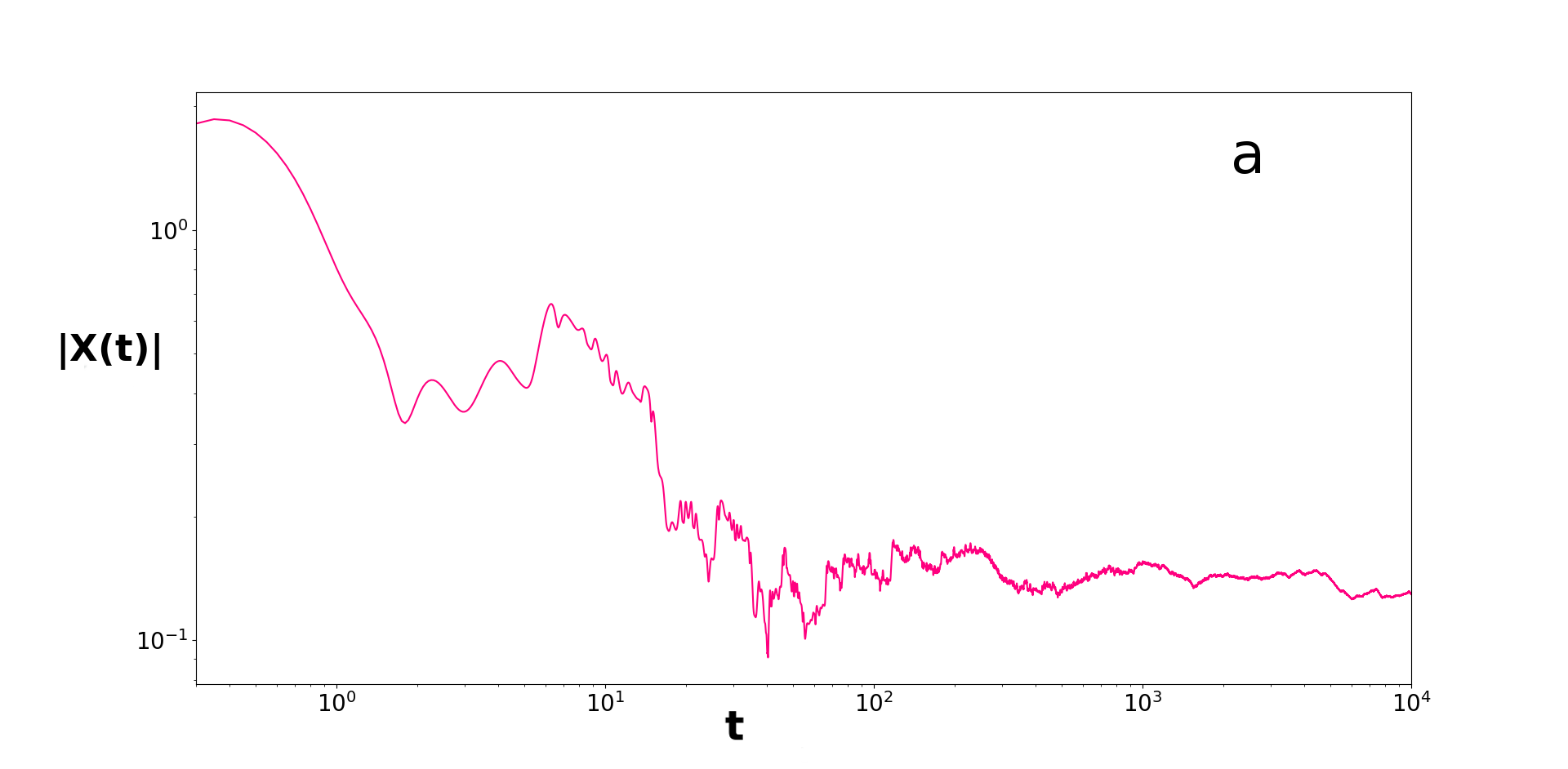}
\includegraphics[width=7.5cm,height=5.5cm]{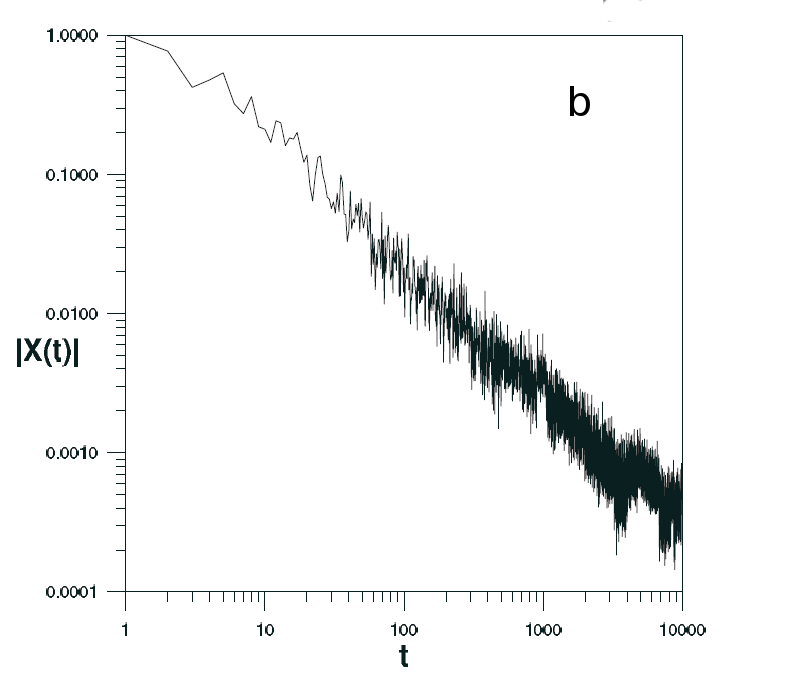}
\caption{The absolute value of the `finite time LCN' (a) for a classical trajectory with initial conditions $x_0=\dot{x}_0=1$ of the system \eqref{Ham} with $\omega=(\sqrt{5}-1)/2$ and  $\varepsilon=3$ (b) for  the corresponding quantum trajectory. In the first case the value of $|\chi|$ stabilizes at $|\chi| \simeq  0.13$, while in the second case $|\chi|$  decreases proportionally to $t^{-1}$, therefore  LCN=0.}\label{lcn}
\end{figure}

\section{NODAL POINTS AND X-POINTS}
The Bohmian equations of motion \eqref{bohmeq} become singular when $\Psi=0$, namely when
\begin{align}
\Re(\Psi)=\Im(\Psi)=0.
\end{align}
The solutions of the above system of equations define the so called `nodal points' of the wavefunction.

\begin{figure}[H]
\centering
\includegraphics[scale=0.3]{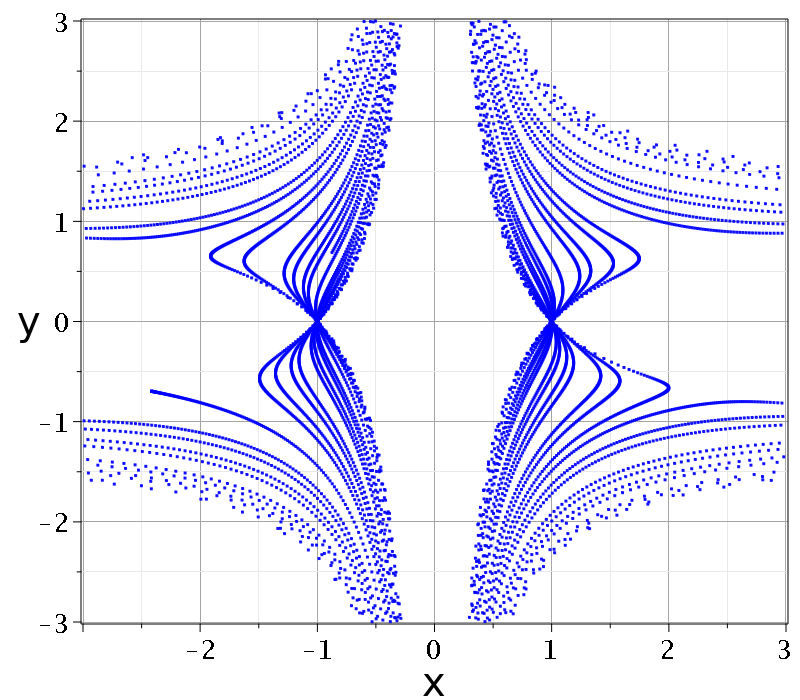}
\caption{The trajectory of the moving nodal point of the system of Fig.~\ref{troxies} (Eqs.~\eqref{nps}).}\label{nl}
\end{figure}

In the particular case of the wavefunction \eqref{sys9} the Bohmian equations of motion  read (for $\omega_1=1$):
\begin{align}
&\frac{dx}{dt}=-\frac{a\sin(t)+b\sqrt{\omega_2}y\sin\Big((1+\omega_2)t\Big)}{G}\\&
\frac{dy}{dt}=-\frac{b\sqrt{\omega_2}x\Big[ax\sin(\omega_2t)+\sin\Big((1+\omega_2)t\Big)\Big]}{G},
\end{align}
where $a=\sqrt{2}c_1, b=2c_2$ and
\begin{align}
G=1+2ax\cos(t)+2b\sqrt{\omega_2}xy\cos\Big((1+\omega_2)t\Big)+a^2x^2+2ab\sqrt{\omega_2}x^2y\cos(\omega_2t)+b\sqrt{\omega_2}x^2y^2
\end{align}
If  $\omega_2$ is irrational then there are both ordered and chaotic trajectories. In particular the nodal point is at 
\begin{align}\label{nps}
x_N=-\frac{\sin\Big((1+\omega_2)t\Big)}{a\sin(\omega_2t)},\quad y_N=-\frac{a\sin(t)}{b\sqrt{\omega_2}\sin\Big((1+\omega_2)t\Big)},\quad
\end{align}
and it moves along the lines of Fig.~\ref{nl}.

\begin{figure}[H]
\centering
\includegraphics[scale=0.32]{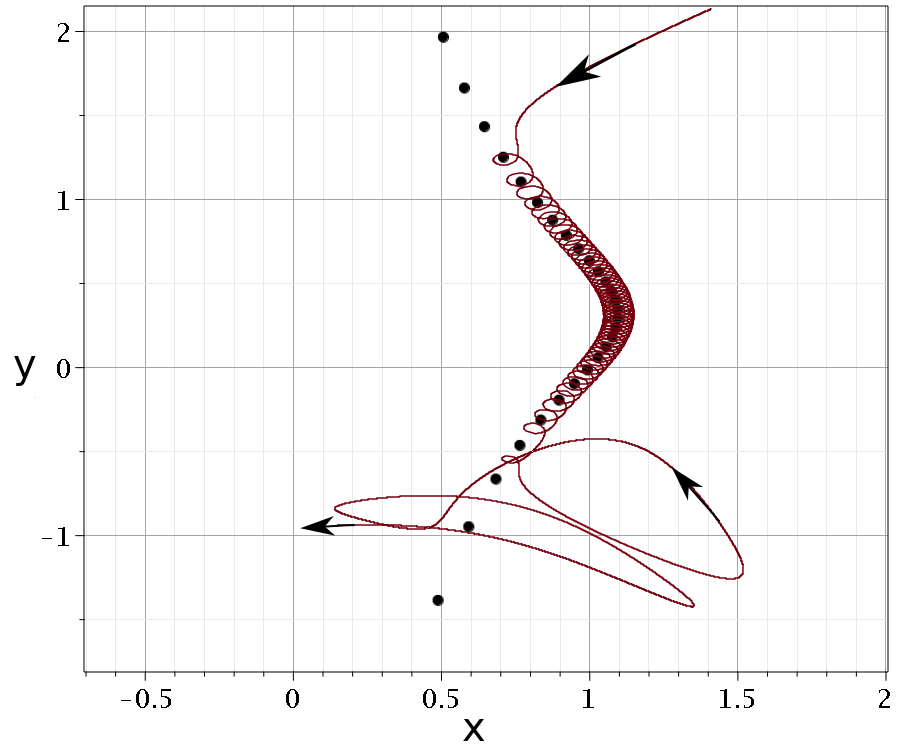}
\caption{A chaotic trajectory that approaches for some time a moving nodal point (black dots) ($x(0)=1.41,y(0)=2.134$).}\label{spira}
\end{figure}

For other solutions of the Schr\"{o}dinger equation we may have 2, 3 etc. nodal points \cite{efth2009} or even an infinity of nodal points \cite{tzemos2019bohmian}. The trajectories of particles very close to the nodal points form  spirals for some time. E.g. in Fig.~\ref{spira} a trajectory is captured close to the nodal point for some time and then escapes away from its neighbourhood. 

\begin{figure}[H]
\centering
\includegraphics[scale=0.5]{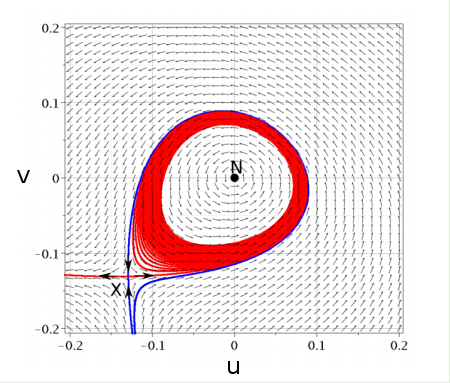}
\caption{The Bohmian flow close to a nodal point $N$. The point $X$ is an unstable stationary point in the system centered at the  nodal point $N$ ($u=x-x_N, v=y-y_N$). We observe the unstable asymptotic curves (red) and stable asymptotic curves (blue). In this figure we show the NPXPC at $t=1.27$ for $\omega_2=\sqrt{2}/2,c_1=1/\sqrt{2}, c_2=1/2$.}\label{xp}
\end{figure}

\begin{figure}[H]
\centering
\includegraphics[scale=0.65]{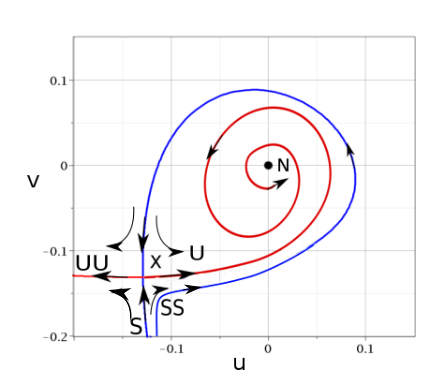}
\caption{The asymptotic curves of the X-point ($U,UU$ unstable and $S, SS$ stable, schematic). Trajectories approaching the X-point close to the stable asymptotic curves are deviated close to the unstable asymptotic curves.}\label{asymptotic}
\end{figure}

\begin{figure}[H]
\centering
\includegraphics[width=8.5cm,height=8cm]{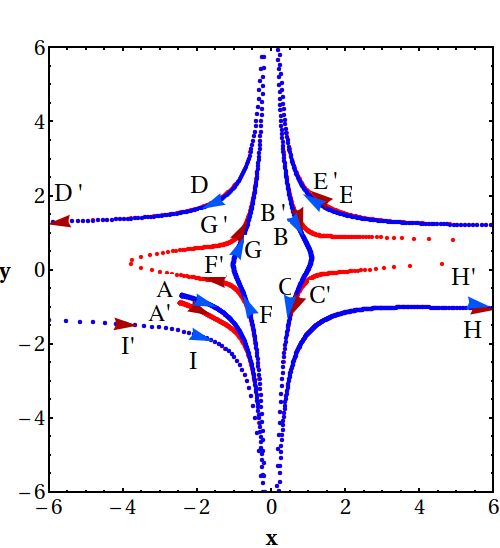}
\caption{The initial arcs of the nodal point (blue) and the X-point (red). The letters refer to the times in parentheses. The unprimed letters correspond to nodal points and the primed letters correspond to the  X-points: $AA'(t = 0), BB'(t =
2.25), CC'(t = 3.45), DD'(t = 4), EE'(t = 5.1), FF'(t = 5.85), GG'(t = 6.75), HH'(t = 8.69),  II'(t = 9.06)$.}\label{arcs}
\end{figure}

The general form of the Bohmian flow close to a moving nodal point is shown in Fig.~\ref{xp}. In a coordinate system centered at a the nodal point there is an unstable hyperbolic point, the `X-point', which has 4 eigenvectors, 2 stable opposite to each other and 2 unstable opposite to each other as well. At the X-point start  4 asympotic curves tangent to these eigenvectors (Fig.~\ref{asymptotic}). One of these asymptotic curves approaches along a spiral the nodal point, while the other 3 asymptotic curves lead far away from the nodal point. 
The X-point moves following the motion of the nodal point. In Fig.~\ref{arcs} we give the initial arcs of the trajectories of the nodal point and the X-point in an inertial frame of reference. We mark certain points along these trajectories at the same times. The nodal point and the X-point form a characteristic geometrical structure of the Bohmian flow, the so called `nodal point-X-point complex' (NPXPC) which is responsible for the production of chaos.


\begin{figure}[hb]
\centering
\includegraphics[scale=0.5]{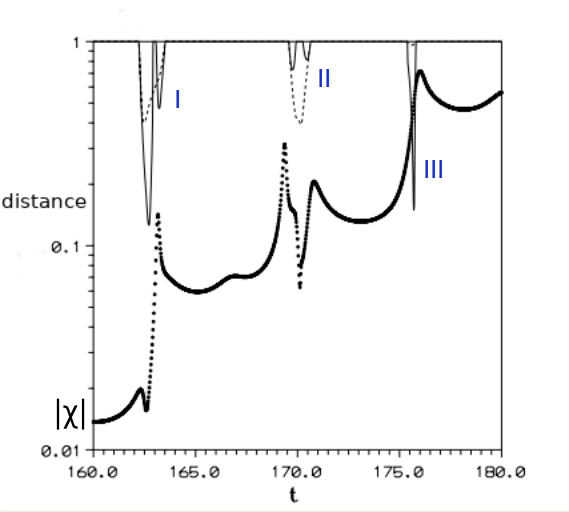}
\caption{The changes of the absolute value of the finite time LCN, $|\chi|$,  when a particle approaches the NPXPC (thick black curve made by bold dots). The thin curve gives the distance of the particle from the nodal point and the dotted curve gives the distance from the X-point. The value of $|\chi|$ increases abruptly during three approaches (I, II and III).}\label{distance} 
\end{figure}

\section{THE ONSET OF CHAOS}

Chaos is introduced when particles approach the neighbourhood of the nodal point and in particular the X-point. A particle can approach the X-point if its trajectory is close to a stable asymptotic curve of the X-point. Then the particle escapes from the  neighbourhood of the X-point  along a trajectory close to one of its unstable asymptotic curves. The cumulative action of the NPXPCs on the trajectories  leads to the emergence of chaos.

In fact during the approach of a particle to the X-point the absolute value of $|\chi|$, in general increases and thus it generates a positive LCN.

The changes of $|\chi|$ become larger as the approaches to the NPXPC become closer (smaller distances $d$ in Fig.~\ref{distance}). Furthermore the effect is larger when the velocity of the particle with respect to the nodal point is small, because then the nodal point can influence the trajectory of the particle for a longer time. In practice, as the particles may have various directions in the inertial plane, this last requirement implies that the velocity $v_N$ of the nodal point in the inertial coordinate system must be small. In fact the deviations $\xi$ are inversely proportional to $dv_N$ (see the detailed discussion of the adiabatic approximation in our paper \cite{efth2009}). However the velocity of the nodal point should not be zero because then the adiabatic approximation cannot be applied.

The topology of Fig.~\ref{asymptotic} changes in time.
Namely in Fig.~\ref{asymptotic} we note three basic characteristics. (a) The asymptotic curve that forms the spiral around the nodal point $N$ is unstable.  (b) The nodal point is an attractor and the asymptotic curve approaches it counterclockwise. c) Furthermore the upper stable asymptotic curve surrounds the spirals around $N$. 

\begin{figure}[H]
\centering
\includegraphics[height=5cm, width=6cm]{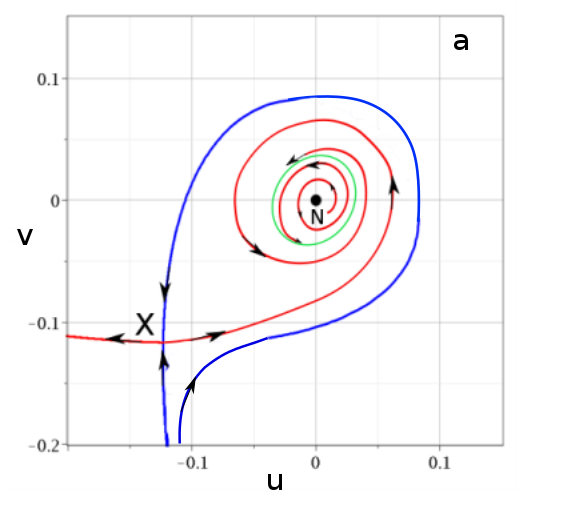}
\includegraphics[height=5cm, width=5.75cm]{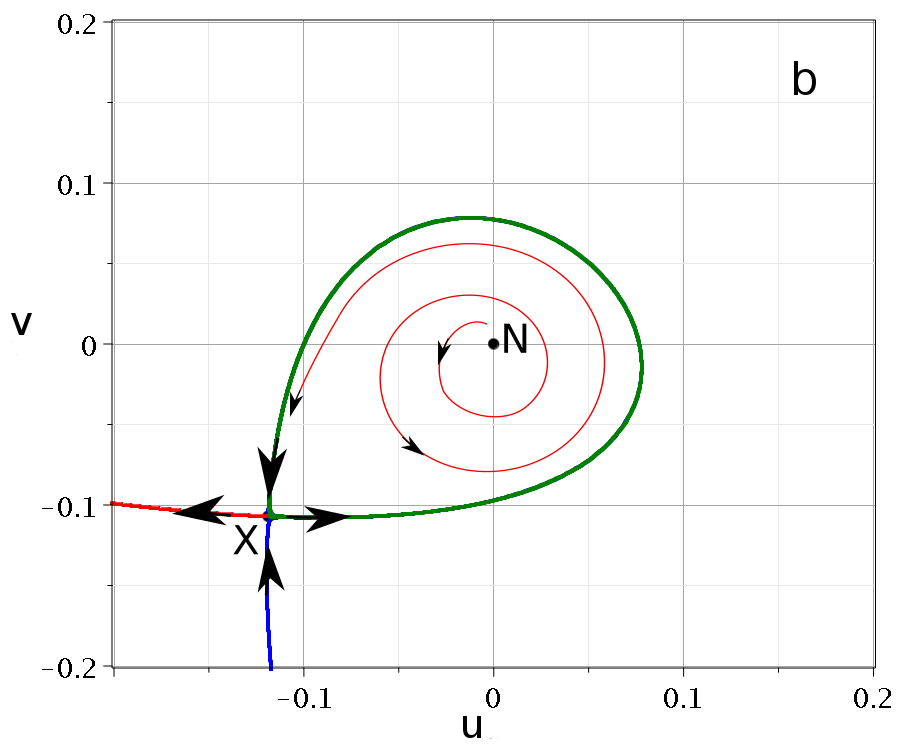}\\
\includegraphics[height=5cm, width=6cm]{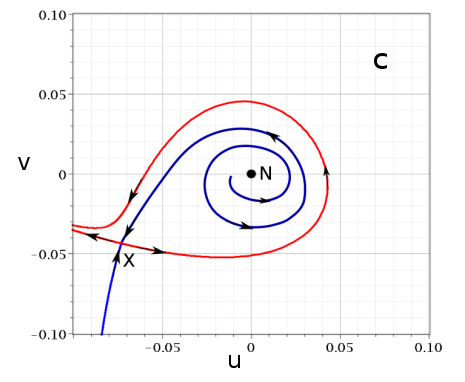}
\caption{Changes of the asymptotic stable (blue) and unstable (red) curves and of the trajectories close to $N$. a) After a Hopf bifurcation from the nodal point $N$, when $N$ becomes a repellor and a limit cycle (green) is formed. b) When the limit cycle reaches the X-point c) After the limit cycle has disappeared the asymptotic curves have  different forms from those of Fig.~\ref{asymptotic}.}\label{hopf}
\end{figure}

As time increases the nodal point becomes a repellor after a Hopf bifurcation and the motions close to it deviate outwards counterclockwise (Fig.~\ref{hopf}a). Then close to the nodal point a limit cycle is formed which is approached by the trajectories around $N$ and by the inner stable (spiral) asymptotic curve from X. As time increases further the limit cycle moves outwards and finally it reaches the X-point (Fig.~\ref{hopf}b). Then the unstable asymptotic curve on the right,  and the stable asymptotic curve (that was previously  surrounding the spirals), are joined together and form a loop that starts from $X$ and ends again at $X$. At a little later time the stable and unstable asymptotic curves are again separated but in a different way (Fig.~\ref{hopf}c). Namely the stable asymptotic cuvre that reaches $X$ from above is a spiral trajectory starting asymptotically from $N$ and the unstable asymptotic curve from the point $X$ that starts moving to the right, surrounds the spirals and continues outwards to the left.

Note that only in Fig.~\ref{asymptotic} the trajectories of particles approaching $X$ between the lower stable asymptotic curves  approach initially the nodal point $N$. However as later the nodal point becomes a repellor these trajectories deviate from the nodal point. This explains why in Fig.~\ref{spira} (which is drawn in the inertial plane) the trajectory of a particle starting away from the nodal point is trapped as a helix around the moving nodal point $N$ but later it is expelled from the neighbourhood of $N$. 

In any case most trajectories  approach from time to time the  NPXPCs and become chaotic but only a few of them are trapped for some time very close to the nodal point forming spirals. On the other hand trajectories that stay far from the NPXPCs for all times are in general ordered.

\section{COMMENSURABLE FREQUENCIES}
\begin{figure}[H]
\centering
\includegraphics[scale=0.2]{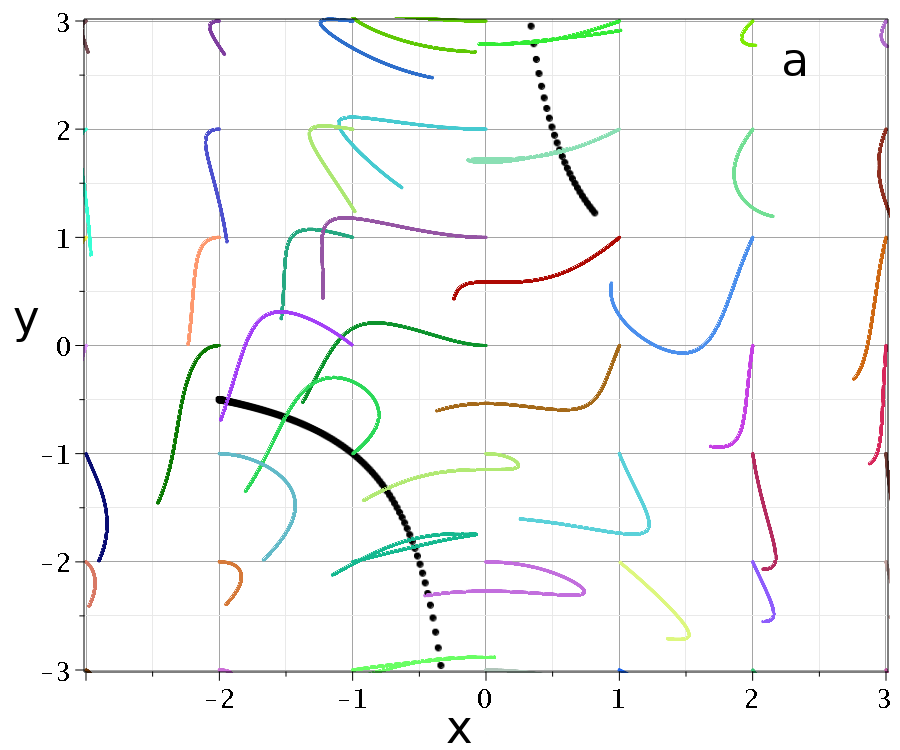}
\includegraphics[scale=0.2]{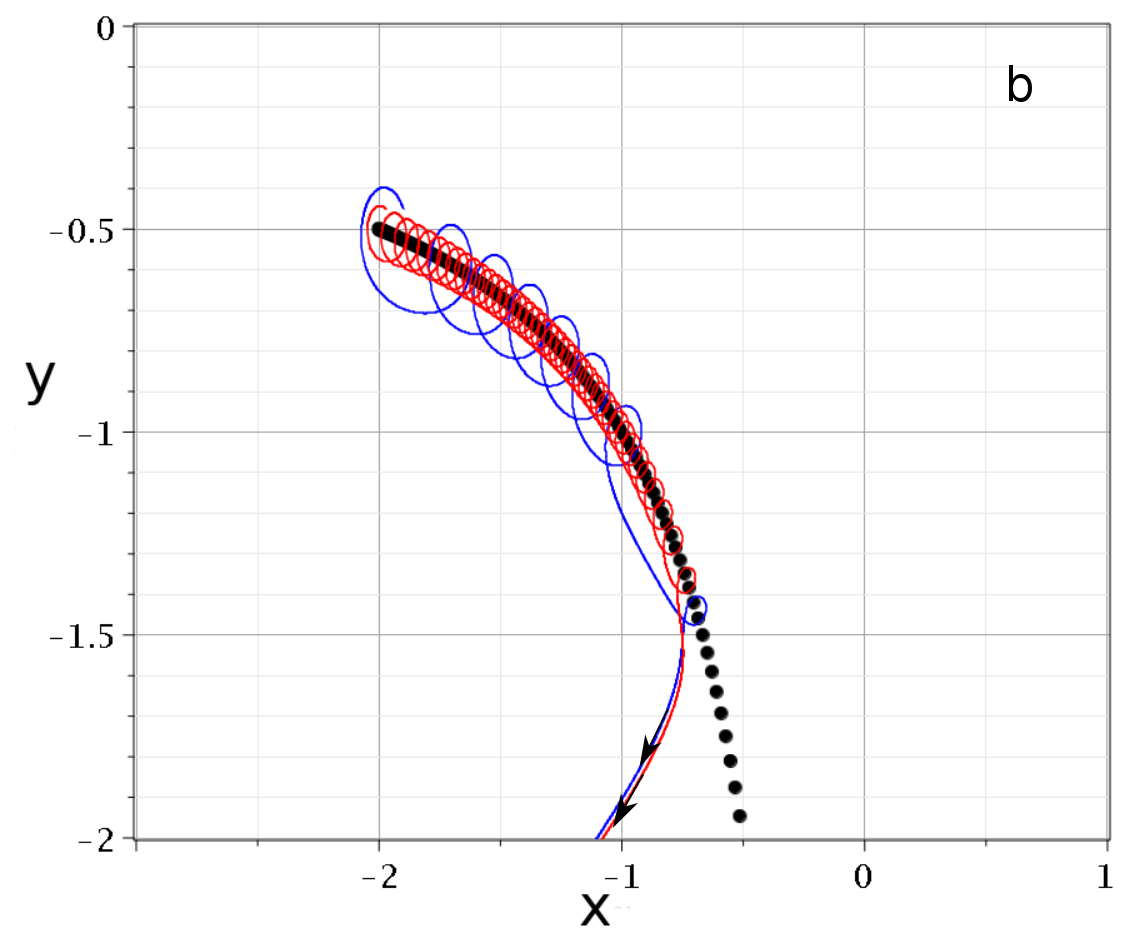}
\caption{Periodic trajectories and the nodal lines in a system with $\omega_1=\omega_2=1$. (a) Several simple trajectories (colored) and two parts of a nodal line (black). (b) Two trajectories starting close to a nodal line form spirals and escape downwards. The smaller their distance from the nodal line, the denser their spiral motion.}\label{periodic}
\end{figure}

\begin{figure}[H]
\centering
\includegraphics[scale=0.45]{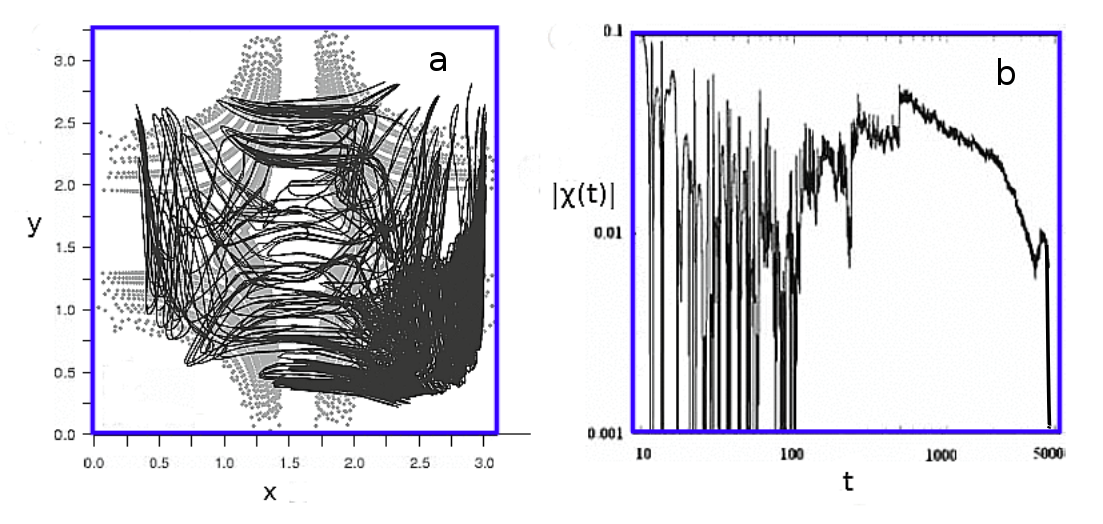}
\caption{A periodic trajectory in the case $\omega_1=1,\omega_2=629/676$. b) The absolute value of the  corresponding finite time LCN, $|\chi(t)|$ up to one period.}\label{periodic2}
\end{figure}

When $\omega_2/\omega_1$ is a rational number, all trajectories are periodic, as it was shown by Wu and Sprung in \cite{wu1999quantum},  therefore there is no chaos. However, while for simple rationals the periodic trajectories are simple as in the case $\omega_1=\omega_2=1$ (Fig.~\ref{periodic}a,b) in cases of high order rationals the periodic trajectories may be quite complicated. Most trajectories are simple (Fig.~\ref{periodic}a) but the trajectories approaching closely the nodal point $N$ make spiral motions around it (Fig.~\ref{periodic}b) for some time. A particular example is shown in Fig.~\ref{periodic2}a  where $\omega_2=629/676$. This particular trajectory was calculated   in the paper \cite{konkel1998regular}  with insufficient accuracy and was considered as chaotic. However, by calculating this trajectory with 50 digit accuracy we could establish that this trajectory is periodic (thus ordered) with a large period $T=2\pi\times 676=4247.4$.

If we calculate the finite time LCN of this trajectory we see in Fig.~\ref{periodic2}b, that $|\chi|$ remains large for a long time and but after  one, two, etc, periods it goes to zero. In the long run $|\chi|$ decreases like $t^{-1}$ beyond the limits of Fig.~\ref{periodic2}b, thus the trajectory is ordered.

Similar phenomena appear whenever $\omega_2/\omega_1$ is a high order rational number. The trajectories may look as chaotic for a long time, but the fact that $\omega_2/\omega_1$ is rational guarantees that all the trajectories are periodic (ordered).

\section{CHAOS IN 3-D SYSTEMS}
\begin{figure}[H]
\centering
\includegraphics[scale=0.3]{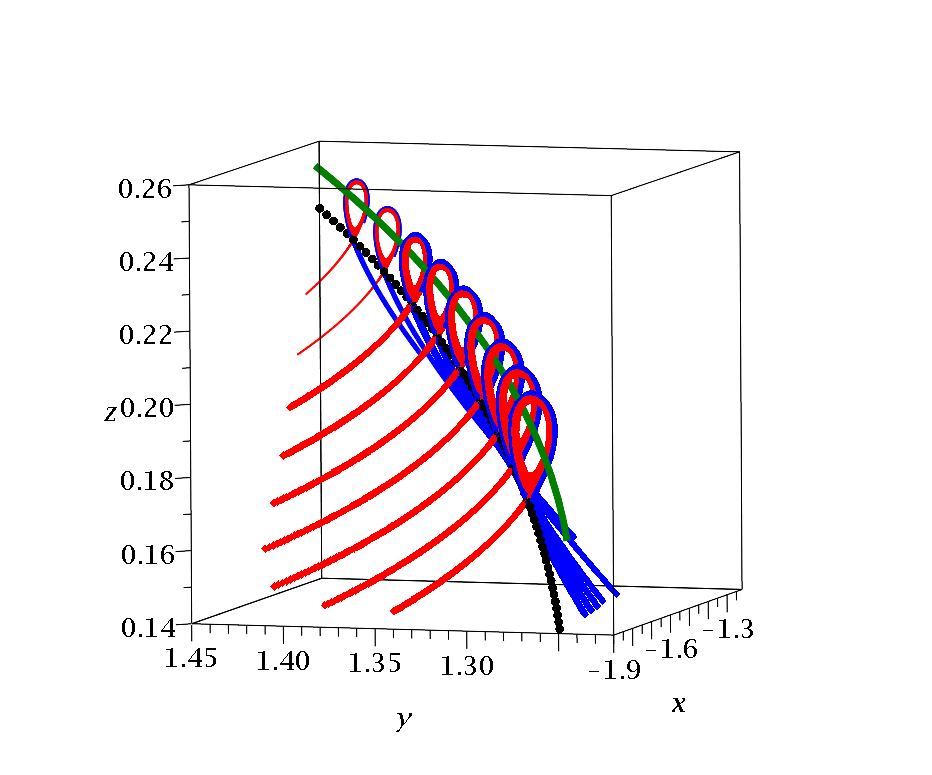}
\caption{A nodal line (green) and an X-line (black) in a 3-d system, together with the asymptotic curves of several X-points  (red and blue). This is the so called `3d structure of nodal and X-point complexes'.}\label{3dxp}
\end{figure}

\begin{figure}[H]
\centering
\includegraphics[scale=0.25]{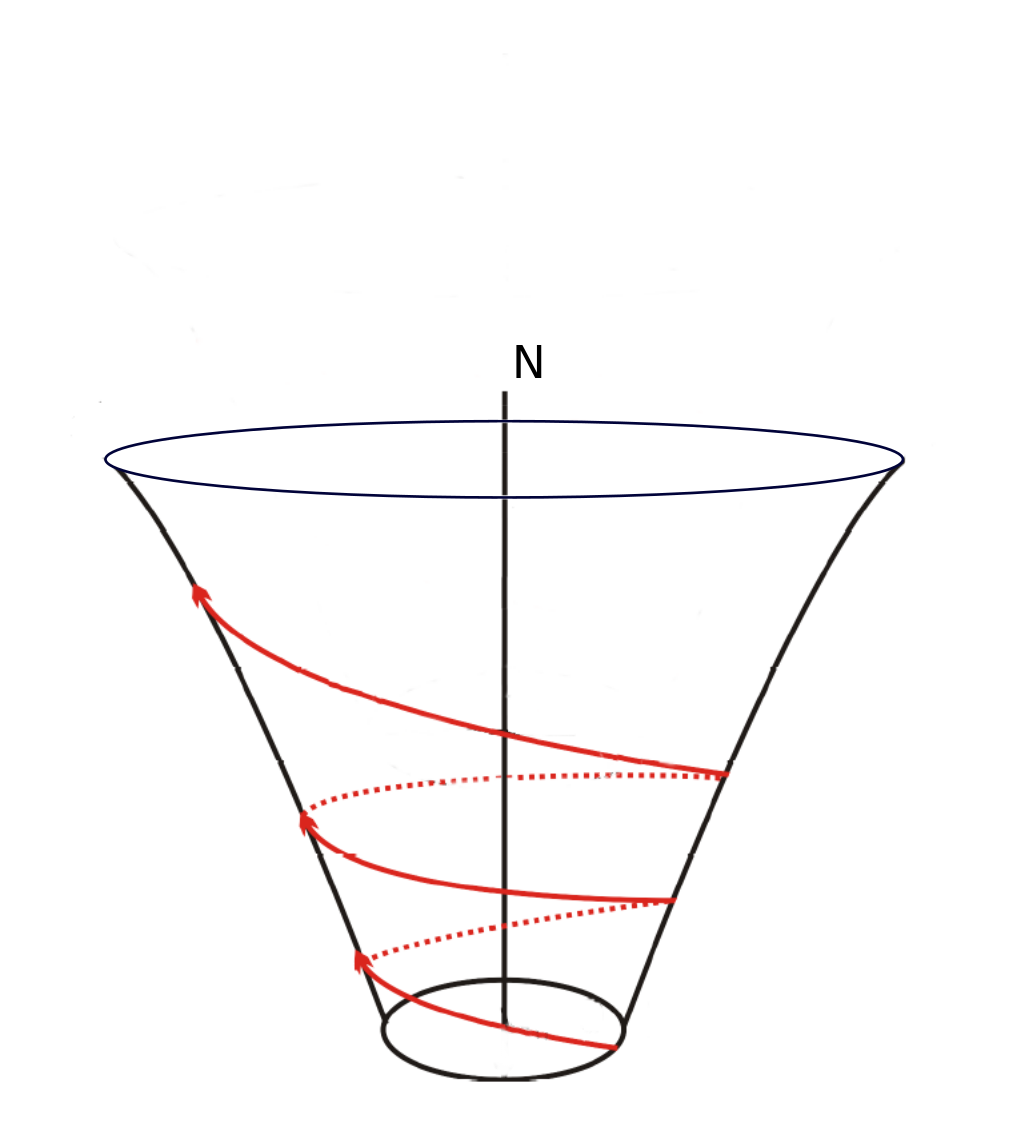}
\caption{A helical trajectory around a moving nodal point in a 3-d system.}\label{helical}
\end{figure}

The general mechanism for the emergence of chaos in the trajectories of an arbitrary 3d Bohmian system has been discussed by Tzemos et al. in \cite{tzemos2018origin}. The main results are the following. In 3 dimensions there are nodal lines along which the wavefunction goes to zero $(\Psi_R=\Psi_I=0)$ (green curve  in Fig.~\ref{3dxp}).
Close to a nodal line there is an `X-line' consisting of unstable equilibria in coordinate frames centered at the nearby nodal points, as shown by the black  curve in Fig.~\ref{3dxp}. The trajectories of particles close to a nodal point are nearby planar spirals on  a plane perpendicular to the nodal line that we call `F-plane', as it was shown by Falsaperla and Fonte \cite{falsaperla2003motion}. However  as the nodal point is moving in 3d space these trajectories are helical (Fig.~\ref{helical}) for some time, in close analogy to the 2d case where we have spiral motion. Chaos is generated whenever a trajectory approaches the region close to the nodal line and the X-line, which now form the so called `3d structures of nodal and X-point complexes'.


\begin{figure}[H]
\centering
\includegraphics[scale=0.4]{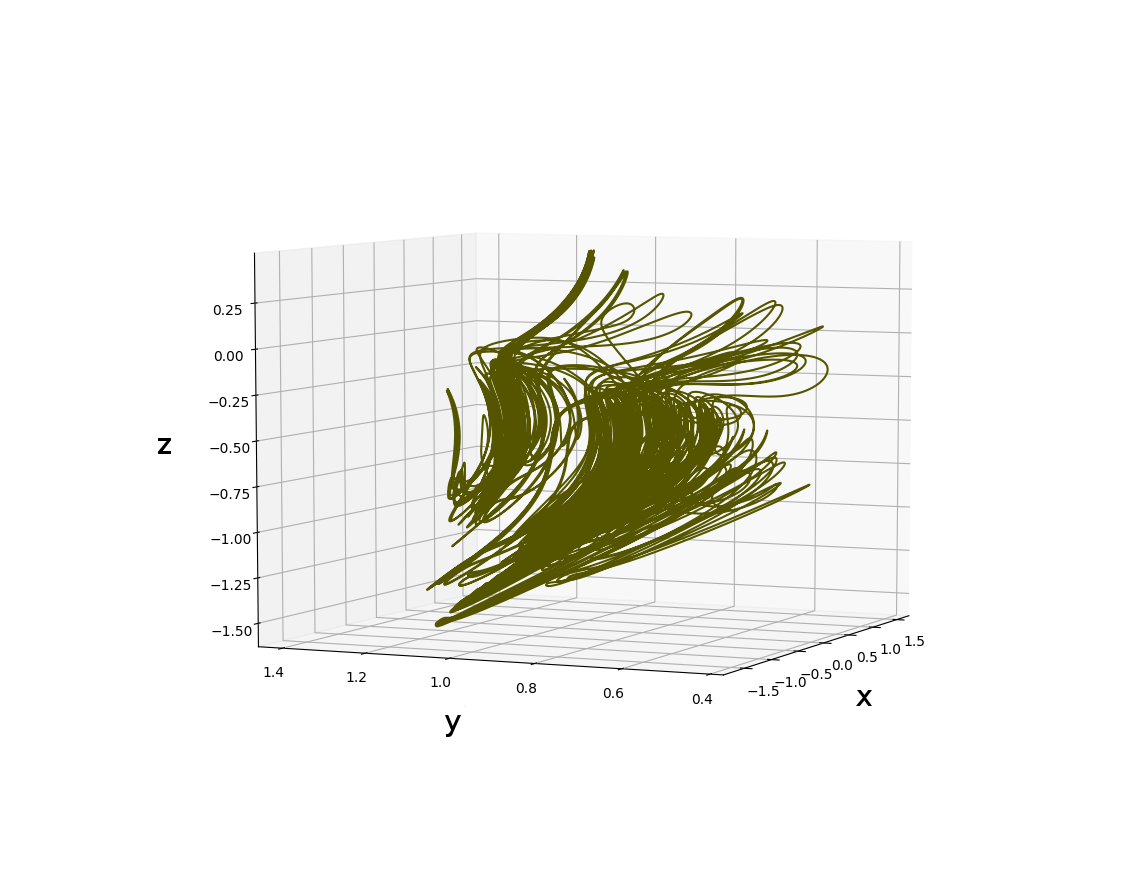}
\caption{A 3d chaotic Bohmian trajectory for $t\in[0,600]$ with initial conditions $x_0=1, y_0=0.7, z_0=-1$.}\label{fullchaos}
\end{figure}

 In general the chaotic trajectories fill a volume in the 3-d space (Fig.~\ref{fullchaos}). However there are cases in which the trajectories lie on 2-surfaces embedded in the 3-d configurational space, due to the existence of an exact integral of motion. This phenomenon is called `partial integrability' \cite{contopoulos2017partial,tzemos2018integrals}.

Such cases appear in systems with 3 eigenfunctions of the form 
\begin{align}\label{psigen}
\Psi=a\Psi_{p_1,p_2,p_3}+b\Psi_{r_1,r_2,r_3}+c\Psi_{s_1,s_2,s_3},
\end{align}
with
\begin{align}\label{eigenstate}
\Psi_{n_1,n_2,n_3}(\vec{x})=\prod_{k=1}^3\frac{\Big(\frac{m_k\omega_k}{\hbar\pi}\Big)^{\frac{1}{4}}\exp\Big(\frac{-m_k\omega_kx_k^2}{2\hbar}\Big)}{\sqrt{2^{n_k}n_k!}}H_{n_k}\Big(\sqrt{\frac{m_k\omega_k}\hbar}x_k\Big),
\end{align}
where $n_1,n_2,n_3$ stand for  the  quantum numbers, $\omega_1,\omega_2,\omega_3$ for their frequencies and $E_1,E_2,E_3$ for their energies. For a given set $(n_1,n_2,n_3)$ we have
\begin{align}
E=\sum_{i=1}^3(n_i+\frac{1}{2})\hbar\omega_i
\end{align}
Partial integrability appears when there are certain relations between the indices $p_i,r_i,s_i,\,\, (i=1,2,3)$ given in \cite{contopoulos2017partial,tzemos2018integrals}. 
For example if the wavefunction is 
\begin{align}
\Psi=a\Psi_{100}+b\Psi_{010}+c\Psi_{002},
\end{align}
with $a=b=c=1/\sqrt{3}$, so that $|a|^2+|b|^2+|c|^2=1$, and $m_k=\hbar=1$, the integral surface is:
\begin{align}\label{peareq}
x^2+y^2+z^2/2-\ln(z)/2\omega_3=C
\end{align}

\begin{figure}[H]
\centering
\includegraphics[scale=0.25]{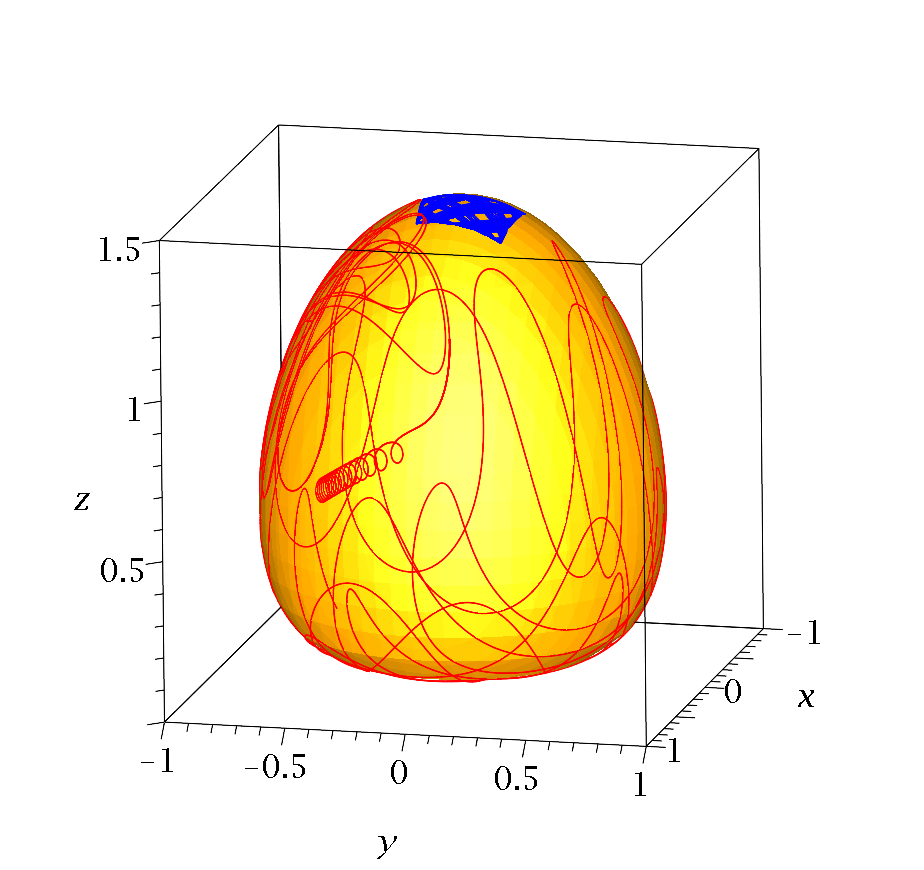}
\caption{An ordered (blue) and a chaotic (red) trajectory on the closed pear-shaped integral surface in the partially integrable case of Eq.~\eqref{peareq}.} \label{pear}
\end{figure}

For a particular value of the constant $C$ and $\omega_3=\sqrt{3}$ we have the surface of Fig.~\ref{pear}. On this surface moves the nodal point and the trajectories of particles starting on it . The trajectories are either ordered or chaotic. In Fig.~\ref{pear} we see an ordered trajectory (blue) that does not approach a nodal point, and a trajectory starting close to a nodal point (red). This trajectory follows for some time the nodal point by forming a helix around it but later it deviates considerably from it and wanders around on the integral surface.

\begin{figure}[H]
\centering
\includegraphics[scale=0.25]{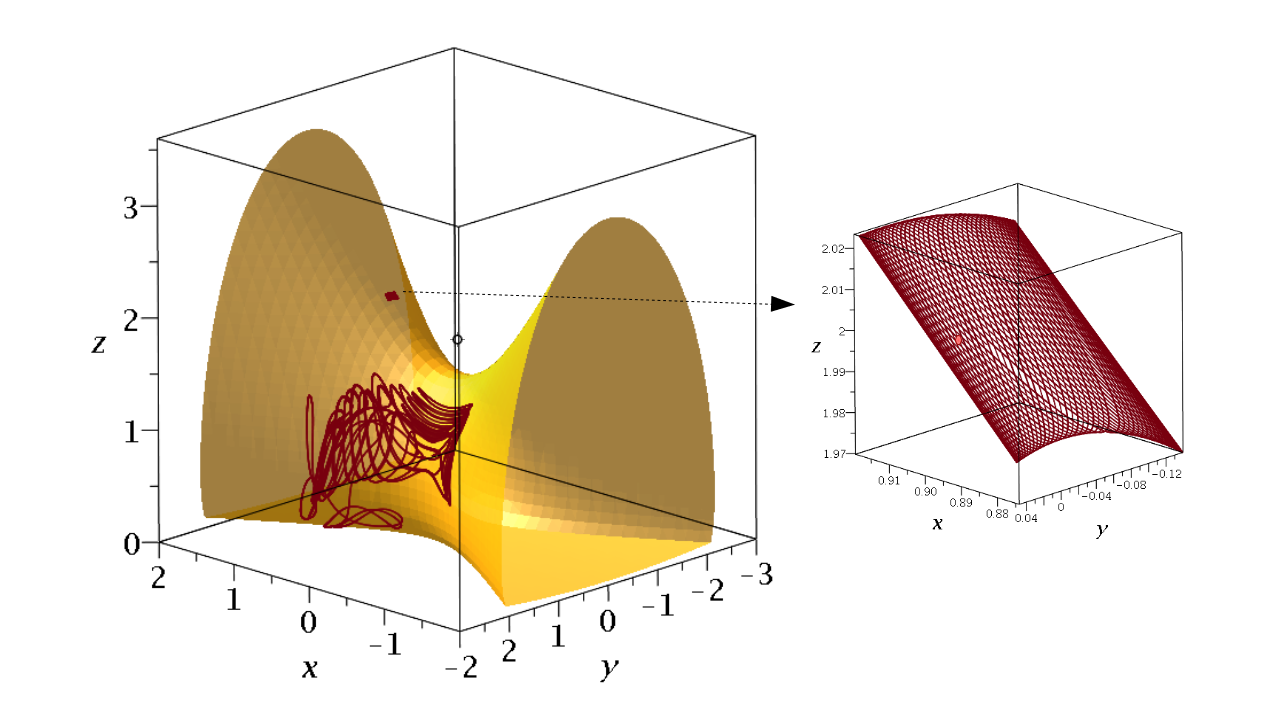}
\caption{An ordered and a chaotic  trajectory on the open integral surface in the partially integrable case of Eq.~\eqref{anoiktieq}. The ordered trajectory is magnified in the right figure.} \label{anoikti}
\end{figure}

The integral surface may be closed, as in Fig.~\ref{pear} or open as in Fig.~\ref{anoikti}. This last case refers to a wavefunction 
\begin{align}
\Psi=a\Psi_{000}+b\Psi_{110}+c\Psi_{102}
\end{align}
in which the trajectories and the nodal lines lie on the open surface  
\begin{align}
-x^2+y^2+z^2/2-\ln(z)/2\omega_3=C\label{anoiktieq}
\end{align}
Finally there are completely integrable cases \cite{tzemos2018integrals}. This happens  if e.g., in  Eq.~\eqref{psigen},  one of the constants, say $c$, is zero. In this case there are two integrals of motion, representing two surfaces in the 3-d space. The trajectories are the intersections of these surfaces, therefore they cannot be chaotic.


As an example we consider the case 
\begin{align}
\Psi=a\Psi_{100}+b\Psi_{001}
\end{align}
In this case there are two integrals of motion
\begin{align}
&x^2+z^2=c_1\\&
y=c_2
\end{align}
Then the trajectories are circles on planes perpendicular to the y-axis (ordered).

\section{CHAOS AND BORN'S RULE IN ENTANGLED SYSTEMS}
According to Born's rule (BR) the probability density  $P$ of finding a quantum particle in a region of space is given by
\begin{align}\label{Born}
P=|\Psi|^2.
\end{align}
BR is one of the cornerstones of QM and has never been doubted by the experiments. However, while in the standard approach of QM the Born rule is a postulate, in BQM it can be considered as an emergent property of quantum systems.

 It is well known that if initially the distribution of the particles satisfies the relation $P_0=|\Psi_0|^2$, then Schr\"{o}dinger's equation leads to the relation \eqref{Born} for all times. If, however, initially $P_0$ is different from $|\Psi_0|^2$, then it is expected that in the long run $P$ will tend to $|\Psi|^2$. This process is called dynamical relaxation towards a quantum equilibrium, which is exactly Born's rule and has been studied by several authors \cite{bohm1954model,  valentini1991signalI,valentini1991signalII, durr1992typicality,   bohm2006undivided, wisniacki2003dynamics, valentini2005dynamical, efthymiopoulos2006chaos, goldstein2007uniqueness, towler2011time, abraham2014long,  tzemos2020chaos}. However, there are cases where there is no such relaxation. E.g. if all the trajectories are ordered then there is no change of the initial distribution $P_0$, thus if  $P_0\neq|\Psi_0|^2$ there is no relaxation of $P$ towards $|\Psi|^2$. Therefore chaos is a necessary condition in order to  approach relaxation.

But this condition is not sufficient. In fact some chaotic cases were found where this relaxation was not satisfied \cite{delis2012quantum}. Therefore a systematic study of chaos and relaxation is required. This was realized in an entangled system of 2 qubits\footnote{In quantum computing we define the qubit as the unit of quantum information. The physical realization of a qubit is a quantum mechanical system with two well defined states \cite{nielsen2004quantum}.} composed of coherent states of 1-d harmonic oscillators.

A 1-d coherent state $|\alpha(t)\rangle$  is defined as the eigenstate of the annihilation operator $\hat{\alpha}$ of the quantum harmonic oscillator
\begin{equation}
\hat{\alpha}|\alpha(t)\rangle=A(t)|\alpha(t)\rangle,\quad A(t)=|A(t)|\exp(i\phi(t)).
\end{equation}  The corresponding wavefunction in the position representation is:
\begin{equation}
\Psi(x,t)=\Bigg(\frac{m\omega}{\pi\hbar}\Bigg)^{\frac{1}{4}}
\exp\Bigg[-\frac{m\omega}{2\hbar}\Bigg(x-\sqrt{\frac{2\hbar}{m\omega}}
\Re[A(t)]\Bigg)^2+i\Bigg(\sqrt{\frac{2m\omega}{\hbar}}
\Im[A(t)]x+\xi(t)\Bigg)\Bigg],
\end{equation}\label{cs}
with
\begin{eqnarray}
\Re[A(t)]=a_0\cos(\sigma-\omega t),
\Im[A(t)]=a_0\sin(\sigma-\omega t)\\ \xi(t)=\frac{1}{2}\Big[a_0^2
\sin(2(\omega t-\sigma))-\omega t\Big],
\end{eqnarray}
where
$\sigma=\phi(0)$ is the initial phase of $A$, $\phi(t)=\phi(0)-\omega t$ and $a_0\equiv |A(0)|$.

We consider now a wavefunction of the form:
\begin{align}\label{qubits}
\Psi&=c_1\Psi_R\Psi_L+c_2\Psi_L\Psi_R
\end{align}
where  $\Psi_R,\Psi_L$ 
are particular solutions of Schr\"{o}dinger's equation depending either on $(x,t)$ or $(y,t)$ defined as 
\begin{eqnarray}
\nonumber \Psi_R(i,t)\equiv \Psi(i,t;\omega=\omega_i,m=m_i,\sigma=\sigma_i),\,i=x,y\\
\Psi_L(i,t)\equiv \Psi(i,t;\omega=\omega_i,m=m_i,\sigma=\sigma_i+\pi), i=x,y
\end{eqnarray}
and $\Psi_R(\Psi_L)$ refer to a one-dimensional coherent state starting at $t=0$ on the right(left) from the center  along a certain axis ($x$ or $y$) and $|c_1|^2+|c_2|^2=1$. We take $m_x=m_y=1, \omega_x=1, \omega_y=\sqrt{3}$ (thus $\omega_x/\omega_y$ is irrational), $\sigma_x=\sigma_y=0$  and $a_0=5/2$.
The value of $c_2$ defines the degree of entanglement. Namely if $c_2=\sqrt{2}/2=0.707$ we have $c_1=c_2$ (maximum entanglement) and if $c_2=0$ we have zero entanglement. The analytical form  of the wavefunction \eqref{qubits} and of its corresponding equations of motion are given in the Appendix.

The nodal points of the wavefunction are given by the equations
\begin{eqnarray}
\label{xnod}&x_{nod}={\frac {\sqrt {2}
\left( k\pi\,\cos \left( 
\omega_{y}\,t \right) +\sin \left( 
\omega_{y}\,t \right) \ln  \left( 
\left| {\frac {c_{1}}{c_{2}}} \right|  
\right)  \right) }{4\sqrt {\omega_{x}}a_{0}\,\sin \left(  
\omega_{xy}  t \right) } }\\&
\label{ynod}y_{nod}={\frac {\sqrt {
2} \left(k\pi\, \cos \left( \omega_{x}t 
\right) +\sin \left( \omega_{x}t \right) 
\ln  \left(  \left| 
{\frac {c_{1}}{c_{2}}} \right|  \right)  
\right) }{4\sqrt {\omega_{y}}a_{0}\,\sin 
\left( \omega_{xy}\,t \right) }}
\end{eqnarray}
with $k\in Z $, $k$ is  even for $c_1c_2<0$
and  odd for $c_1c_2>0$, and $\omega_{xy}\equiv \omega_x-\omega_y$. 
We choose to work with positive $c_1,c_2$, Therefore the  $k's$ are odd  and infinite in number.  

\begin{figure}[hb]
\centering
\includegraphics[scale=0.4]{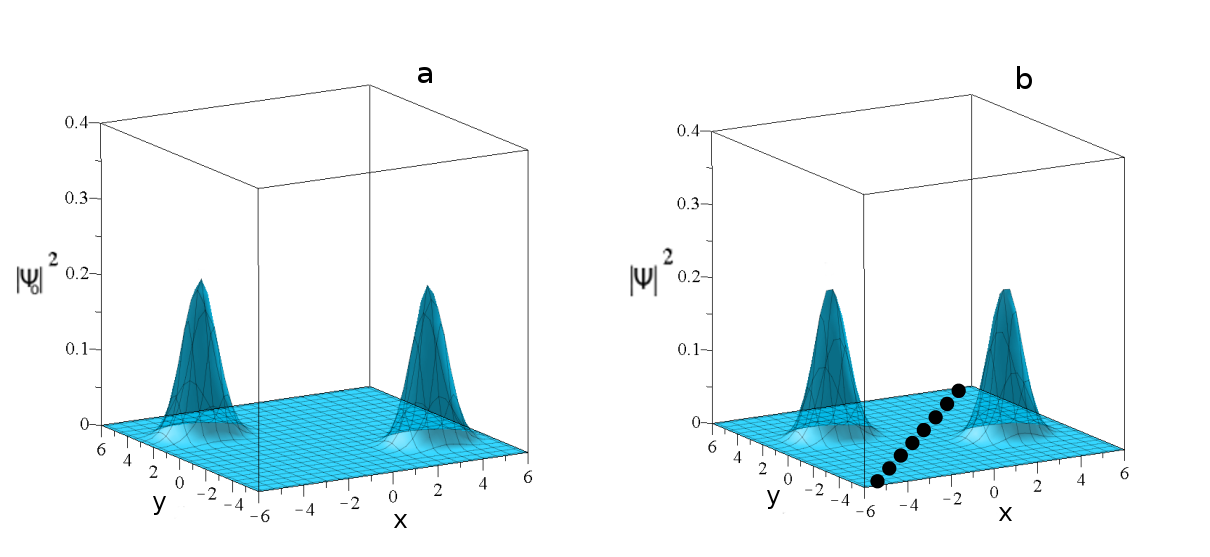}
\caption{The form of $|\Psi|^2$ in the case of maximal entanglement at $t=0$ (a) and at $t\simeq 0.53$ (b). At $t=0$ the nodal points are at infinity, while at $t\simeq 0.53$ the nodal points (black dots) are at finite distances. } \label{katanomes}
\end{figure}

\begin{figure}[H]
\centering
\includegraphics[scale=0.35]{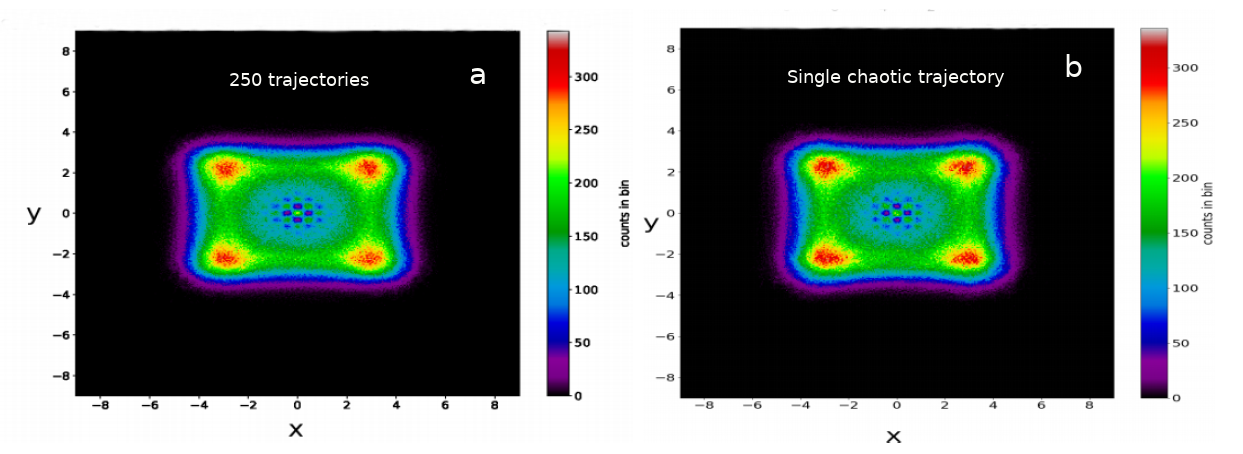}
\caption{Colormaps for the occupation number of the cells of a 360x360 grid on the xy plane for: a) the distribution of the points of 250 trajectories in the case of maximal entanglement $c_2=\sqrt{2}/2$ (the sampling time of the points is $\Delta t=0.05$) up to $t=1000$ and b) the distribution of the points of one trajectory (with $x_0=2, y_0=-2$) up to a time $t=250000$. They are practically the same.} \label{bell_katanomi}
\end{figure}

In the case of maximum entanglement,  $c_1=c_2=\sqrt{2}/2$, all the trajectories are chaotic. In this case Born's rule gives initially  two blobs of particles around the points $x_0=\pm 3.54, y_0=\mp 2.69$ (Fig.~\ref{katanomes}a). At time $t=0$ all the nodal points are at infinity. A little later, however, the nodal points come closer to the origin $(0,0)$ (Fig.~\ref{katanomes}b). Then  we collect the points along every trajectory at every $\Delta t=0.05$ and find after a long time the distributions of the trajectory points shown by colors in Fig.~\ref{bell_katanomi}. The limiting distribution of  250 trajectories satisfying initially Born's rule $P_0=|\Psi_0|^2$ is given in Fig.~\ref{bell_katanomi}a, while in Fig.~\ref{bell_katanomi}b we show the limiting distribution of a single trajectory for $t=250000$ (in both cases we kept the sampling time fixed and equal to $\Delta t=0.05$). We observe that Figs.~\ref{bell_katanomi}a, \ref{bell_katanomi}b are almost the same. This was found to be true for every initial condition. Consequently all  chaotic trajectories are  ergodic as they all give the same distribution after a long time. Thus in this case any initial distribution of particles will always lead to Born's rule.

\begin{figure}[H]
\centering
\includegraphics[scale=0.4]{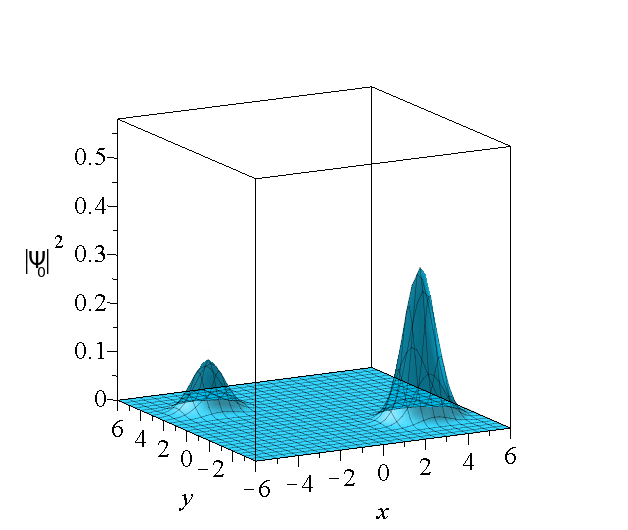}
\caption{The form of $|\Psi_0|^2$ in the case $c_2=0.5$ at $t=0$. We observe the unequal heights of the two blobs. Now there some initial conditions close to the center of the large blob that  produce ordered trajectories.} \label{05katanomiI}
\end{figure}

If now we decrease $c_2$ from its maximum value $c_2=\sqrt{2}/2$ the Born distribution consists of two unequal blobs (Fig.~\ref{05katanomiI}) the larger one around the point $x_0=3.54, y_0=-2.69$ and the smaller one around $x_0=-3.54, y_0=2.69$.

\begin{figure}[H]
\centering
\includegraphics[scale=0.28]{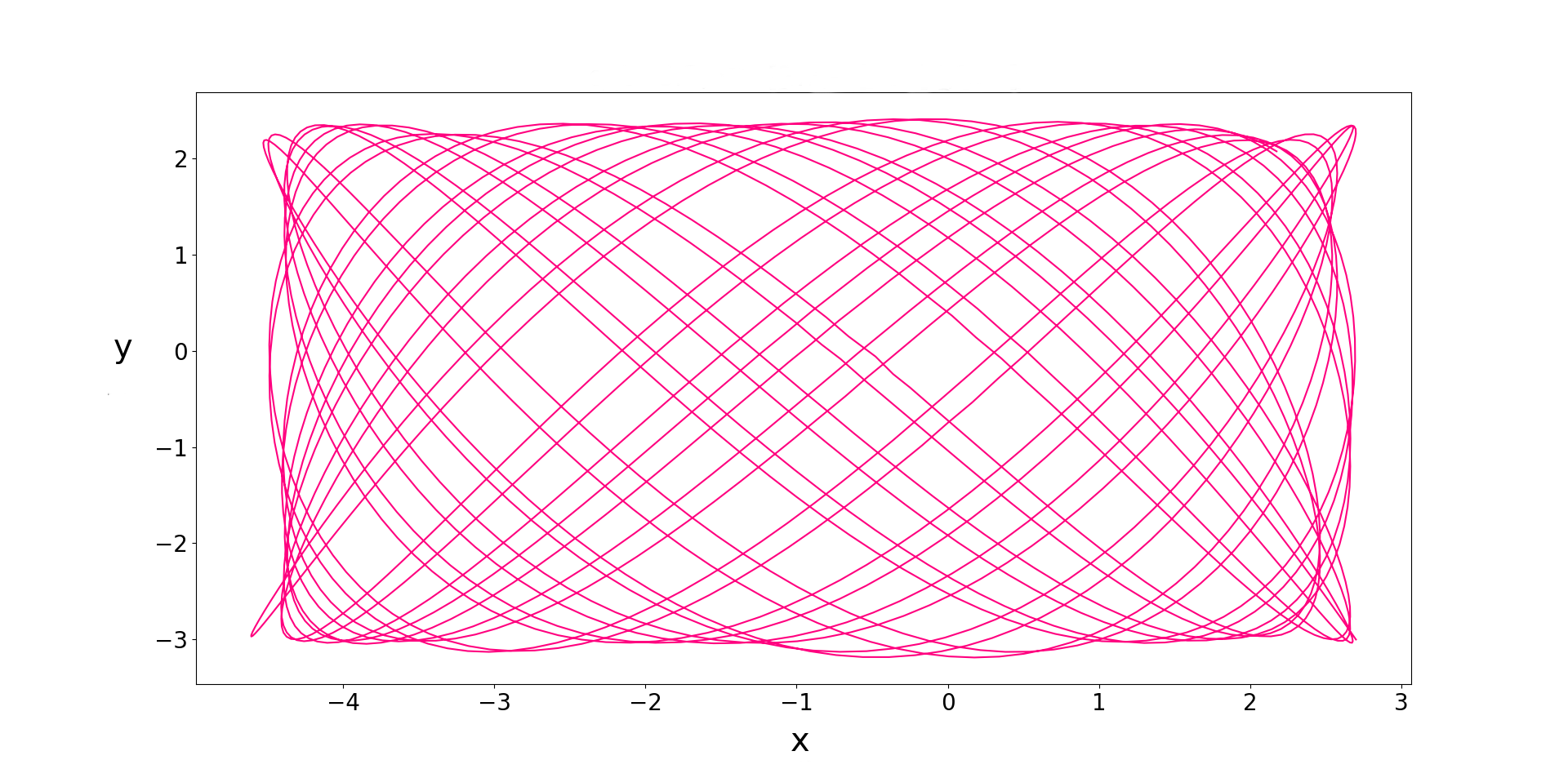}
\caption{An ordered trajectory in the case $c_2=0.2$  ($x_0=2.7,y_0=-3$ and $t\in[0,200]$) has the form of a distrorted Lissajous figure.} \label{troxia02}
\end{figure}

\begin{figure}[H]
\centering
\includegraphics[scale=0.3]{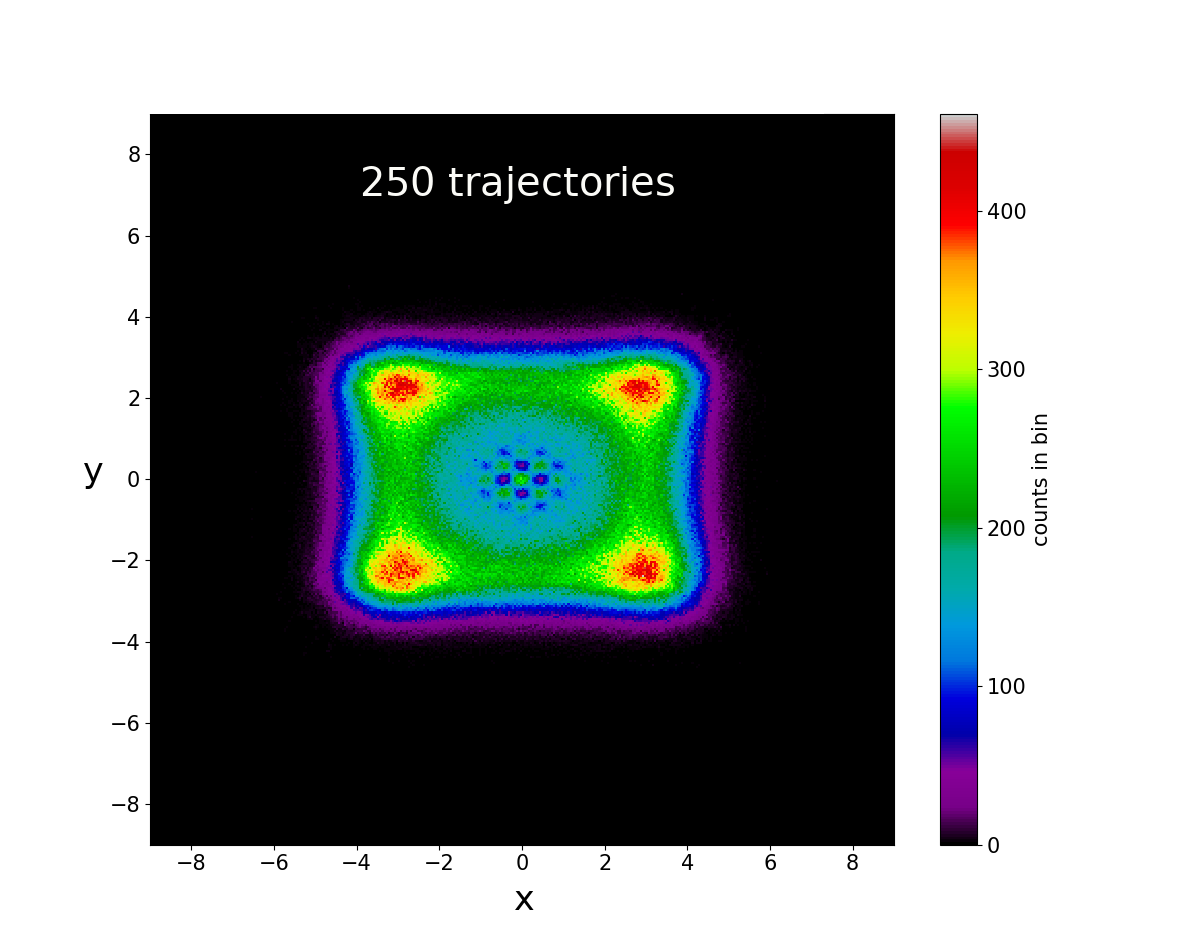}
\caption{The final distribution of the points of 250 trajectories in the case $c_2=0.5$. It is the same with the final distribution of any single chaotic trajectory.} \label{05katanomiII}
\end{figure}

In these cases a few trajectories close to the center of the large blob are ordered. These have the shape of deformed Lissajous figures (Fig.~\ref{troxia02}) and they are obviously not ergodic. However if their number is small they do not affect appreciably the distribution of points that leads to Born's rule. E.g. this is the case of $c_2=0.5$. In this case  the final distribution is the same for most trajectories and satisfies approximately the final distibution of the set of trajectories following the Born rule (Fig.~\ref{05katanomiII}).

\begin{figure}[H]
\centering
\includegraphics[scale=0.65]{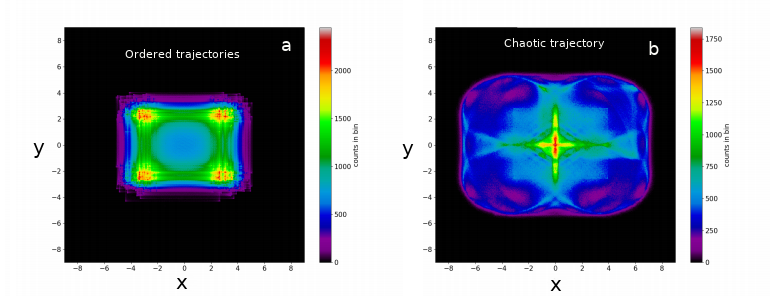}
\caption{a)The final distribution of the points of 250 trajectories in the case $c_2=0.001$. In this case most trajectories satisfying initially Born's rule are ordered (deformed Lissajous figures). b) The final distribution of the points of a chaotic trajectory. All chaotic trajectories give the same final distribution (thus they are ergodic), but this distribution is very different from the final distribution of the ordered trajectories of (a).} \label{0001katanomi}
\end{figure}

However if $c_2=0.2$ most trajectories of the initial Born's distribution $P_0=|\Psi_0|^2$ are ordered and not chaotic (neither ergodic). The same is true for other small values of $c_2$, as e.g. in the case $c_2=0.001$. Then the  final distribution of the points of the  trajectories satisfying initially Born's rule consists of a superposition of many deformed Lissajous figures (Fig.~\ref{0001katanomi}a). On the other hand most individual trajectories, either within the blobs of the initial Born distribution, or further away from them are chaotic and ergodic, and they tend to form a different, in general, final distribution (Fig.~\ref{0001katanomi}b). Thus if we take any collection of chaotic trajectories their final distribution (Fig.~\ref{0001katanomi}b) is quite different from that of Born's rule.

Finally in the limit of no entanglement ($c_2=0$) all trajectories are ordered (exact Lissajous figures) and of course they are not chaotic, nor ergodic. Therefore unless the initial distribution follows Born's rule, any other initial distribution gives a different final distribution, and Born's rule is not satisfied.

We conclude that regardless of  the degree of entanglement 
the chaotic trajectories are ergodic. In the case of strong entanglement an arbitrary initial distribution will tend to that of Born's rule after some time. On the other hand if the entanglement is weak the limiting distribution of chaotic trajectories is in general different from that provided by Born's rule, which is dominated by ordered trajectories. However there are arguments \cite{bohm1954model} that even in these cases the Born distribution is established after very long times. If this is correct it would require extremely long times, well beyond the limits of our numerical experiments. 

\section{CONCLUSIONS}
In this  review  we presented without many mathematical details the aspects  of BQM that we studied in the last 15 years which are:
\begin{enumerate}
\item The development of the mechanism responsible for the production of chaos in arbitrary 2-d and 3-d Bohmian systems.
\item The comparative study between classically chaotic/ordered
systems and their quantum analogues within the framework of BQM.
\item The coexistence of order and chaos in most Bohmian systems in 2 and 3 dimensions.
\item The existence of certain 3d systems which are partially integrable, where
ordered and chaotic trajectories coexist on certain integral surfaces given by
analytical integrals of motion.
\item The interplay between chaos and entanglement in systems of theoretical and technological importance.
\item The effect of chaos and entanglement on the dynamical approximation to Born's rule in BQM.
\end{enumerate}

Our results raise new questions 
for future studies such as:
\begin{enumerate}
\item How entanglement affects the evolution of multipartite Bohmian systems.
\item How chaos affects the dynamics of open quantum systems.
\end{enumerate}

\section{APPENDIX}
The analytical formula of the wavefunction \eqref{qubits} in the position
representation reads:
\begin{align}\label{analytical}
\nonumber\Psi&=c_1\Psi_R\Psi_L+c_2\Psi_L\Psi_R\\&\nonumber=
 \frac{(\omega_{x}\omega_{y})^{1/4}}{\sqrt{\pi}} \exp\left\{ \frac{i}{2}\left[{a_0}^2 (\sin(2{\omega_x}t)+\sin(2{\omega_y}t))-({\omega_x}+{\omega_y})t
\right]\right\}\\&\nonumber\times \exp\Big\{-\frac{1}{2}(\omega_xx^2+\omega_yy^2)-a_0^2[\cos^2(\omega_xt)+\cos^2(\omega_yt)]\Big\}
\\&\times
\Big\{\exp(f_x-f_y-i(g_x-g_y))c_1+\exp(-f_x+f_y+i(g_x-g_y)c_2\Big\}
\end{align}

where 
\begin{align}\label{f}
&f_x=\sqrt{2\omega_x}{a_0}x\cos(\omega_x t), \quad f_y= \sqrt{2\omega_y}{a_0}y\cos(\omega_y t)\\
&g_x=\sqrt{2\omega_x}{a_0}x\sin(\omega_x t), \quad g_y= \sqrt{2\omega_y}{a_0}y\sin(\omega_y t) 
\end{align}

The Bohmian equations of motion produced by the wavefunction \eqref{analytical}
are 
\begin{equation}\label{eqmotion_newx}
\frac{dx}{dt}=-\frac{\sqrt{2\omega_x}{a_0}\left[A\cos(\omega_x t)+B\sin(\omega_x t) \right]}{G} 
\end{equation}

\begin{equation}\label{eqmotion_newy}
\frac{dy}{dt}=\frac{\sqrt{2\omega_y}{a_0}\left[A\cos(\omega_y t)+B\sin(\omega_y t) \right]}{G}
\end{equation}
where
\begin{equation}\label{eqmotion_A}
A=2{c_1}{c_2}{e^{2f_x +2f_y }}\sin(2(g_x-g_y)), \quad B={c_1}^2 {e^{4f_x}}-{c_2}^2{e^{4f_y}}
\end{equation}
and 
\begin{equation}\label{eqmotion_G}
G={c_1}^2 {e^{4f_x+4f_y}}+2{c_1}{c_2}{e^{2f_x +2f_y }}\cos(2(g_x+ g_y)) +{c_2}^2
\end{equation}

\section*{ACKNOWLEDGEMENTS}
This work was supported by the Research Committee of the Academy of Athens.

\section*{References}
\bibliographystyle{iopart-num}
\bibliography{bibliography}

\end{document}